\documentclass[a4paper,10pt]{article}
\usepackage[utf8]{inputenc}

\usepackage{amsmath}
\usepackage{amssymb}
\usepackage{hyperref}
\usepackage{graphicx}
\usepackage{verbatim}
\usepackage{geometry}
\usepackage{tikz}
\tikzstyle{every picture}=[level distance = 8mm, baseline=-0.5ex]
\tikzstyle{prop}=[shape=circle,minimum size=6mm, draw=black!80, fill=green!30]

\renewcommand{\d}{\text{d}}

\newcommand{\simall}[2]{\underset{#1\rightarrow#2}{\sim}}

\begin{document}

\title{A {S}chwinger--{D}yson Equation in the {B}orel Plane: singularities of the solution.}
\author{Marc~P.~Bellon${}^{1,2}$, Pierre~J.~Clavier${}^{1,2}$\\
\normalsize \it ${}^1$ Sorbonne Universités, UPMC Univ Paris 06, UMR 7589, LPTHE, 75005, Paris, France\\
\normalsize \it $^2$ CNRS, UMR 7589, LPTHE, 75005, Paris, France }

\date{}

\maketitle

\begin{abstract}
We map the Schwinger--Dyson equation and the renormalization group equation for the massless Wess--Zumino model in the Borel plane, where the product of functions gets mapped to a convolution product. The two-point function can be expressed as a superposition of  general powers of the external momentum. The singularities of the anomalous dimension are shown to lie on the real line in the Borel plane and to be linked to the singularities of the Mellin transform of the one-loop graph. 
This new approach allows us to enlarge the reach of previous studies
on the expansions around those singularities.
The asymptotic behavior at infinity of the Borel transform of the solution is beyond the reach of analytical methods and we do a preliminary numerical study, aiming to show that it should remain bounded.
\end{abstract}

\textbf{Mathematics Subjects Classification:} 81Q40, 81T16, 40G10.


\section*{Introduction}


The perturbative formulation of quantum field theory (QFT) allows one to compute, order by order, precise quantum effects. Using the Feynman rules, one usually computes a set of diagrams representing 
all possible processes starting and ending with the states we are interested in. To reach more precise results, one has to compute more diagrams. The successes of this approach are beyond count, 
but the most famous of them are QED, electroweak theory and QCD.

However, the number of diagrams to compute grows very quickly with the order of the perturbation theory. Moreover, the diagrams also become more challenging to evaluate, due to the larger number of counterterms to use or the possible 
complications in their topologies. Hence the highest computed order of perturbation theory has grown rather slowly over the past few decades. Let us also notice that there are situations of great 
interest for physicists where perturbation theory breaks down, due to a large coupling constant. The archetypal example of such a situation is the low-energy QCD.

The Schwinger--Dyson equations are a way to reach non-perturbative information on a QFT. Other trails are, for example, lattice QCD or effective models. Schwinger--Dyson equations have 
been applied quite successfully to low-energy QCD. For example in \cite{MeMoRoSa14} they were used to construct a Generalized Parton Distribution satisfying both the theoretical constraints of 
polynomiality, time-reversal invariance and charge conjugation, and the experimental data of Jefferson Lab.

Nonetheless, results coming from Schwinger--Dyson equations for physical systems heavily rely on numerical analysis, and few analytical results are known. The only exact known solutions are 
for linear cases \cite{BrKr99}, \cite{Cl14}. Nevertheless, even a perturbative solution of a Schwinger--Dyson equation carries non-perturbative information and can lead to a better understanding of 
analytical non-perturbative aspects of the theory. In this paper, a step is made in this direction for a non-linear Schwinger--Dyson equation.

In the work \cite{KrYe2006} it was made clear that the renormalization group equation could be used to get the dressed propagator from the anomalous dimension, that could itself be then extracted 
from the Schwinger--Dyson equation. This trail was put into practice in \cite{Be10a} and \cite{BeSc12} for the massless Wess--Zumino model. Extending this approach allowed us to reach perturbative corrections 
to the asymptotic behavior of the anomalous dimension in \cite{BeCl13}.

Notwithstanding its successes, the analysis of \cite{BeCl13} involves symbols of unclear meaning, in a redundant description of the perturbative series involving diverging series. An expansion for the Mellin transform was also used, that could not be 
proved to be exact. These features are quite unsatisfactory and call for a more rigorous analysis. Indeed, a better understanding of the method of \cite{BeCl13} is needed before its use becomes possible in 
more physically relevant models. We will use the Borel transform to understand our divergent series as markers of the simplest singularities of the Borel transform. In this formalism, we will no longer need the expansion of the Mellin transform in terms of pole contributions.

The Borel transform, seen as a morphism of the ring of formal series, allows the definition of the sum of some series of null radius of convergence. Many series of physical interest 
(such as perturbative expansions) are within this class of series. Singularities of the Borel transform introduce differences between Borel sums of the theory in different sectors and give unavoidable non-perturbative contributions. The fundamental mathematical work on the question has been done by Jean \'Ecalle~\cite{Ecalle81}, which coined the word resurgence. Physicists, starting from the works of Manfred Stingl for quantum field theories~\cite{St95,St02}, mostly took home the message that special expansions, dubbed transseries, involving the perturbatively zero quantities \(e^{-B/g}\) could give better approximation for finite value of the coupling~\(g\) than a simple perturbation expansion. The approach is becoming increasingly popular, with many recent publications in theoretical physics having some of the key words of this theory in their titles or abstracts, a few of them are \cite{ChDoDuUn13,ChDoUn14,CoEdScVo14}.
What set apart this work is that the singularities of the Borel transform are studied uniquely through the use of alien calculus, without any reference to saddle points in functional integration or a previous knowledge of the higher order of the perturbative series.

This paper is divided into four parts. The first one is a recall of the methods and results developed in \cite{Be10a}, \cite{BeSc12} and \cite{BeCl13} to study the anomalous dimension of the 
massless Wess--Zumino model. In the second part, we apply the Borel transform to the Schwinger--Dyson equation and the renormalization group equation of this model and write them in a convenient 
form to study the singularities of the Borel transform of the anomalous dimension. The third section is devoted to the study of the singularities of the Borel transform of the anomalous 
dimension. Their localization is found and their transcendental contents are shown to be only odd zetas. The weights of those zetas are also studied. Finally, in the fourth section, 
we perform a numerical and asymptotic analysis of those equations.

\section{Set-up and previous results}

\subsection{The problem}

We work with the massless Wess--Zumino model. The equations we are going to deal with are the renormalization group equation and a minimal Schwinger--Dyson equation for the model. First, 
the model being massless allows us to expand the two-point function in power of the logarithm of the impulsion $L=\ln(p^2/\mu^2)$%
:
\begin{equation} \label{2ptB}
 G(L) = 1 + \sum_{k=1}^{+\infty}\gamma_k\frac{L^k}{k!}
\end{equation}
with $\gamma_1:=\gamma$ the anomalous dimension of the theory. The $\gamma_n$'s are themselves functions of the fine structure constant of the theory, written $a$. The two-point function obeys the 
renormalization group equation:
\begin{equation} \label{renorm_G}
 \partial_L G(a,L) =( \gamma + \beta a\partial_a )G(a,L),
\end{equation}
with $\beta$ the beta function of the theory. Now, the Callan--Symanzik equation leads to $\beta=3\gamma$ (see \cite{Piguet} for a proof of this result) and this renormalization group equation 
gives a recurrence relation for the $\gamma_n$'s:
\begin{equation} \label{renorm_gamma_old}
 \gamma_{k+1} = \gamma(1+ 3a\partial_a)\gamma_k.
\end{equation}
This result was detailed in the thesis \cite{Ye08} and the article \cite{BeSc08} and means that $\gamma$ is enough to fully know $G$.

We will write the Schwinger--Dyson equation as an equation for $\gamma$. Let us take the Schwinger--Dyson equation of the propagator truncated to the first non-trivial term:
\begin{equation}\label{SDnlin}
\left(
\tikz \node[prop]{} child[grow=east] child[grow=west];
\right)^{-1} = 1 - a \;\;
\begin{tikzpicture}[level distance = 5mm, node distance= 10mm,baseline=(x.base)]
 \node (upnode) [style=prop]{};
 \node (downnode) [below of=upnode,style=prop]{}; 
 \draw (upnode) to[out=180,in=180]   
 	node[name=x,coordinate,midway] {} (downnode);
\draw	(x)	child[grow=west] ;
\draw (upnode) to[out=0,in=0] 
 	node[name=y,coordinate,midway] {} (downnode) ;
\draw	(y) child[grow=east]  ;
\end{tikzpicture}.
\end{equation}
In the massless Wess--Zumino model, it is the only important Schwinger--Dyson equation for the computation of renormalization group function. It is also the simplest non-linear Schwinger--Dyson 
equation. Now, the full propagator can be written as the free propagator times the two-point function.
\begin{equation}
 P(p^2) = \frac{1}{p^2}\left(1 + \sum_{k=1}^{+\infty}\gamma_k\frac{L^k}{k!}\right)
\end{equation}
To compute the loop integral, we take its double Mellin transform, so that the logarithms can be recovered from derivations in Mellin parameters:
\begin{equation*}
 \left(\ln p^2\right)^k = \left.\frac{\d^k}{\d x^k}\left(p^2\right)^x\right|_{x=0}.
\end{equation*}
In order to end up with an equation on $\gamma$ only, we take the derivative of (\ref{SDnlin}) with respect to $L$ and set $L$ to zero, which rid us of the divergence. Then we end up with an equation for $\gamma$:
\begin{equation} \label{SDE_old}
 \gamma = a\left.\left(1+\sum_{n=1}^{+\infty}\frac{\gamma_n}{n!}\frac{\text{d}^n}{\text{dx}^n}\right)\left(1+\sum_{m=1}^{+\infty}\frac{\gamma_m}{m!}\frac{\text{d}^m}{\text{dy}^m}\right)H(x,y)\right|_{x=y=0}
\end{equation}
with $H$ the  Mellin transform of the one loop integral:
\begin{equation} \label{def_H}
 H(x,y) = \frac{\Gamma(1-x-y)\Gamma(1+x)\Gamma(1+y)}{\Gamma(2+x+y)\Gamma(1-x)\Gamma(1-y)}.
\end{equation}

\subsection{Asymptotic solution}

Looking directly for an asymptotic solution for the equation (\ref{SDE_old}) is unpractical due to the quadratic growth of the number of terms contributing to a given order. In \cite{Be10a}, it was proposed 
to approximate the Mellin transform $H(x,y)$ by its poles times a suitable analytic extension of their residues.

$H(x,y)$ has poles at $x;y=-k$, $k\in\mathbb{N}^*$ (those poles come from IR divergences) and at $x+y=+k$, $k\in\mathbb{N}$ (from UV divergences). Both kind of poles arise when a subgraph 
becomes scale invariant for some value of the Mellin variables.

Expanding the IR poles
\begin{equation}
 \frac{1}{k+x} = \frac{1}{k}\sum_{n=0}^{+\infty}\left(-\frac{x}{k}\right)^n,
\end{equation}
shows that the contribution $F_k$ of such a pole has the form
\begin{equation} \label{form_F}
 F_k = \frac{1}{k}\left(1+\sum_{n=1}^{+\infty}\left(-\frac{1}{k}\right)^n\gamma_n\right).
\end{equation}
The renormalization group equation (\ref{renorm_gamma_old}) then gives
\begin{equation} \label{equa_F}
 \gamma(1 + 3a\partial_a) F_k = -k F_k + 1.
\end{equation}
For the UV poles, one has to take care of the numerators. Let $N_k(\partial_{L_1},\partial_{L_2})$ be the numerator of the contribution $L_k$ of the pole at $x+y=k$. Then 
$N_k(\partial_{L_1},\partial_{L_2})=Q_k(\partial_{L_1}\partial_{L_2})$, with $Q_k(xy)$ the suitable expansion of the residue of $H$ at $x+y=k$. Then, as shown in \cite{Be10a}, $L_k$ obeys:
\begin{equation} \label{equa_H}
 (k-2\gamma - \beta a\partial_a)L_k = N_k(\partial_{L_1},\partial_{L_2})G(L_1)G(L_2)|_{L_1=L_2=0}.
\end{equation}
In~\cite{Be10a}, the function $H(x,y)$ was approximated by its first poles at $x=-1$, $y=-1$ and $x+y=+1$, giving the following approximating function in~(\ref{SDE_old}):
\begin{equation} \label{appr}
 h(x,y) = (1+xy)\left(\frac{1}{1+x} + \frac{1}{1+y}-1\right) + \frac{1}{2}\frac{xy}{1-x-y} + \frac{1}{2}xy.
\end{equation}
This means that we only use the contributions $F\equiv F_1$ of the poles $1/(1+x)$ and $1/(1+y)$ and $L\equiv L_1$ of the pole $xy/(1-x-y)$ to compute \(\gamma\). Then the 
renormalization group equations (\ref{equa_F}) and~(\ref{equa_H}) for $F$ and $L$ and the Schwinger--Dyson equation~(\ref{SDE_old}) with the approximate function $h(x,y)$ defined in
 (\ref{appr}), give the three coupled non-linear differential equations:
\begin{equation}
 \begin{cases}
  & F = 1 - \gamma(3a\partial_a+1)F , \\
  & L = \gamma^2 + \gamma(3a\partial_a+2)L  ,\\
  & \gamma = 2a F -a -2a\gamma( F-1) + \frac{1}{2}a(L-\gamma^2) .
 \end{cases}
\end{equation}
To get an asymptotic solution, we expand $F$, $L$ and $\gamma$ in power of $a$: $F=\sum f_na^n$, $L=\sum l_na^n$ and $\gamma=\sum c_n a^n$. Making the assumption that the sequences $\{f_n\}$, $\{l_n\}$ and $\{c_n\}$ have a fast growth, only a few terms in the sums defining the coefficients of a product are dominant and we get three coupled recursions, the solution of which has simple asymptotic properties. All in all, we end up with the two dominant terms in each series:
\begin{align} \label{result_simple}
 \begin{cases}
  & f_{n+1} \simeq -(3n+5) f_n, \\
  & l_{n+1} \simeq 3n l_n ,\\
  & c_{n+1} \simeq -(3n+2) c_n.
 \end{cases}
\end{align}
This recursions nicely fit the numerical results of \cite{BeSc08}.

\subsection{Higher order corrections}

Computing the $1/n$ corrections to (\ref{result_simple}) is quickly tedious. So in \cite{BeCl13} we defined two formal series $A=\sum A_n a^n$ and $B=\sum B_n a^n$ satisfying exactly the asymptotic relation (\ref{result_simple}):
\begin{align}
 \begin{cases}
  & A_{n+1} = -(3n+5)A_n \\
  & B_{n+1} = 3nB_n.
 \end{cases}
\end{align}
The symbols corresponding to the formal series $A$ and $B$ obey (up to some finite polynomial in $a$) to the following differential equations
\begin{align} \label{relation_symboles}
 \begin{cases}
  & 3a^2\partial_a A = -A-5aA \\
  & 3a^2\partial_a B = B.
 \end{cases}
\end{align}
Then, our strategy was to expand $F_K$, $L_k$ and $\gamma$ using those symbols, which encode the asymptotic properties of the solution:
\begin{align}
 \begin{cases}
  & F_k = f_k +Ag_k+Bh_k \\
  & L = l_k+Am_k+Bn_k \\
  & \gamma = a(c+Ad+Be)
 \end{cases}
\end{align}
with $f_k$, $l_k$, \dots, unknown functions of $a$. Then, we inserted this ansatz into the renormalization group  (\ref{renorm_gamma_old}) and Schwinger--Dyson equations (\ref{SDE_old}) with 
$H$ written as a sum over its poles, replaced the derivatives of $A$ and $B$ using (\ref{relation_symboles}), ignored every mixed terms $AB$, $A^2$, etc., and asked every remaining terms to 
separately vanish. Hence, within this formalism, we obtained three equations for each of the previous ones and solved them order by order in $a$.

This procedure, although quite natural, was not fully justified. We will show here that dropping the mixed terms and asking for every remaining terms to vanish is strictly equivalent to working 
in the vicinity of a singularity of the Borel transform of $\gamma$.

The procedure detailed above allowed us to compute the corrections to the asymptotic solution (\ref{result_simple}) up to the order $a^5$ of $\gamma$ in \cite{BeCl13}. Unexpected cancellations 
of zetas were observed in the solution, so that the weights of the coefficients were lower than expected. We will see here that this effect can be better understood in the Borel plane.

\section{Mapping to the Borel plane}

\subsection{Generalities on the Borel transform}

There are many introductions to the Borel transform, and we do not intend to make a new one. We will only say some useful facts and follow the presentation of \cite{Bo11}.

The Borel transform might be seen as a ring morphism between two rings of formal series:
\begin{eqnarray}
             \mathcal{B}: a\mathbb{C}[[a]] & \longrightarrow & \mathbb{C}[[\xi]] \\
\tilde{f}(a) = a\sum_{n=0}^{+\infty}c_na^n & \longrightarrow & \hat{f}(\xi) = \sum_{n=0}^{+\infty}\frac{c_n}{n!}\xi^n \nonumber
\end{eqnarray}
The idea is that even if $\tilde{f}$ is a purely formal series (that is, has a null radius of convergence), $\hat{f}$ might be convergent. There is an inverse Borel transform (the Laplace 
transform), which matches the usual sum whenever $\tilde{f}$ is convergent, and can give a sense to the sum of divergent series. However, this resummation has to be done in sectors of the 
complex plane, bounded by the lines of singularities of the Borel transform. One speaks of sectorial resummation.  When one crosses such a line of singularities of the Borel transform between two different sectors, the result of the summation changes. This is known as the Stokes 
phenomenon and methods have been devised to compute these changes, and their study is very active, especially in the field of dynamical systems.

The essential properties of the Borel transform that we will use are: first, it is a linear transformation. Secondly, the Borel transform of a point-wise product of functions is the convolution product of the Borel transforms:
\begin{eqnarray} \label{prodB}
 \mathcal{B}(\tilde{f}\tilde{g})(\xi) & = & \hat{f}\star\hat{g}(\xi) \\
                                      & = & \int_0^{\xi}\hat{f}(\xi-\eta)\hat{g}(\eta)\d\eta. \nonumber
\end{eqnarray}
The last line being well defined if and only if $\hat{f}$ and $\hat{g}$ have analytic continuations along a suitable path between \(0\) and~ \(\xi\). A consequence of this relation is that the Borel transform of $a.f$ is the primitive of $\hat{f}$.
\begin{equation} \label{primB}
 \mathcal{B}(a.f)(\xi) = \int_0^{\xi}\hat{f}(\eta)\d\eta
\end{equation}
Another very useful relation, which can easily be proved by manipulating formal series is
\begin{equation} \label{derivB}
 \mathcal{B}\left(a\partial_a\tilde{f}(a)\right)(\xi) = \partial_{\xi}\left(\xi\hat{f}(\xi)\right).
\end{equation}
Finally, we will refer in the following to the plane of $a$ as the physical plane, and the plane of $\xi$ as the Borel plane.

\subsection{The Renormalization Group equation}

We will consider the propagator term without its constant term \(\tilde G= G -1 \)\footnote{One can define the Borel transform of a constant as the formal identity of the convolution product: the 
Dirac $\delta$ ``function''. Since we want to deal only with analytic quantities, we rather choose to omit the $1$ in the Borel transform.}.
Using the relation $\beta=3\gamma$, coming from the 
Callan--Symanzik equation of the massless Wess--Zumino term in (\ref{renorm_G}) leads to the following renormalization group equation for $\tilde{G}$:
\begin{equation}
 \partial_L \tilde{G}(a,L) = \gamma\left(1 + 3a\partial_a\right)\tilde{G}(a,L)+\gamma.
\end{equation}
This equation is easily mapped into the Borel plane by using the rules (\ref{prodB}) and (\ref{derivB}) since $\tilde{G}$ has no constant part and has therefore its Borel transform well defined.
\begin{equation*}
 \partial_L\hat{G}(\xi,L) = \hat{\gamma}(\xi) + \int_0^{\xi}\hat{\gamma}(\xi-\eta)\hat{G}(\eta,L)\d\eta + 3\int_0^{\xi}\hat{\gamma}(\xi-\eta)\partial_{\eta}\left(\eta\hat{G}(\eta,L)\right)\d\eta.
\end{equation*}
Treating the convolution product as a perturbation in this equation, one obtains terms which are proportional to \(L^n\). However, the resultant power series in \(L\) is not really informative and is not suitable for a study of the singularities of the Borel transform. Also, due  to the presence of the derivative with respect to \(\xi\) of \(\hat G\), one cannot expect to find \(\hat G\) as a fixed point.

Integrating by parts the last integral and using $\hat{\gamma}(0)=1$ leads to an equation which will prove itself much more convenient.
\begin{equation}
 \partial_L\hat{G}(\xi,L) - 3\xi\;\hat{G}(\xi,L) = \hat{\gamma}(\xi) + \int_0^{\xi}\hat{\gamma}(\xi-\eta)\hat{G}(\eta,L)\d\eta + 3\int_0^{\xi}\hat{\gamma}'(\xi-\eta)\eta\hat{G}(\eta,L)\d\eta
\end{equation}
Here, if we neglect the convolution parts, we have the order zero solution
\begin{equation}
\hat G(\xi,L) = \frac 1 {3\xi} \hat\gamma(\xi)( e^{3\xi L} - 1),
\end{equation}
using the condition \(\hat G(\xi,0)=0\). Introducing this order zero solution in the convolution products suggests that \(\hat G\) for fixed Borel parameter \(\xi\) can be represented as a superposition of exponentials of \(L\) with parameters between \(0\) and \(3\xi\). Since
\(L\) is the logarithm of \(p^2\), it means that we simply have a general power of the impulsion squared.  However, we would like to have a representation which does not depend on the path joining \(0\) and \(\xi\)
and which easily deals with the singularities we expect to have at the ends of the path, since the order 0 solution has Dirac masses at these points.

We therefore parametrize $\hat{G}$ as a contour integral,
\begin{equation} \label{param_G}
 \hat{G}(\xi,L) = \oint_{\mathcal{C}_{\xi}}\frac{f(\xi,\zeta)}{\zeta}e^{3\zeta L}\d\zeta
\end{equation}
with $\mathcal{C}_{\xi}$ any contour enclosing $0$ and $\xi$. On a contour minimally including the endpoints,  the jump of \(f\) along a cut from \(0\) to \(\xi\) gives a smooth integral, while the singularities at the end points will contribute to singular terms.  The condition that \(\hat G(\xi,0)\) is zero is also easily obtained in this formalism, since the exponential becomes 1 for \(L=0\) and the contour can be expanded to infinity.  It is therefore sufficient that \(f\) have limit 0 at infinity. The renormalization group equation for $\hat{G}$ becomes an equation on $f$, since one can use the same contour for the computation of \(\hat G\) for all the necessary values of \(\eta\) and then, switching the order of the contour integral and the other operations, one can write everything 
as a contour integral on a common path. The \(L\) independent term can also be given the same form, noting 
\begin{equation*}
 1 = \oint_{\mathcal{C}_{\xi}}e^{3\zeta L}\frac{\d\zeta}{\zeta}.
\end{equation*}
(A factor \(1/(2\pi i)\) has been included in the definition of the contour integral \(\oint\) to simplify notations.) One ends up with the following equation for $f$:
\begin{equation} \label{renorm_f}
 3(\zeta-\xi)f(\xi,\zeta) = \hat{\gamma}(\xi) + \int_0^{\xi}\hat{\gamma}(\xi-\eta)f(\eta,\zeta)\d\eta + 3\int_0^{\xi}\hat{\gamma}'(\xi-\eta)\eta f(\eta,\zeta)\d\eta.
\end{equation}
We will see later that this equation is the right one to study the singularities of $\hat{\gamma}$.

\subsection{The Schwinger--Dyson equation}

We start with the Schwinger--Dyson equation in the physical plane (\ref{SDnlin}). In fact we only need its derivative with respect to \(L\) at the renormalization point, which defines \(\gamma\),
\begin{equation}
 \gamma(a) = \left.-a\frac{\partial}{\partial L}\int\d^4qP\left(q^2\right)P\left((p-q)^2\right)\right|_{L=0}
\end{equation}
with $P$ the fully renormalized propagator:
\begin{equation}
 P(p^2) = \frac{1}{p^2}\left(1+\tilde{G}(L(p^2))\right).
\end{equation}
In the following, we will denote simply by $\partial_L$ the operator taking the partial derivative with respect to $L$ and evaluating to $0$. The integral naturally splits in three parts, according to the number of \(\tilde G\) factors,
\begin{equation}
 \gamma(a) = -a\partial_L\left[I_1(L)+2I_2(L)+I_3(L)\right],
\end{equation}
with:
\begin{eqnarray*}
 I_1(L) & = & \int\d^4q\frac{1}{q^2(p-q)^2} + S_1(\mu^2) \\
 I_2(L) & = & \int\d^4q\frac{\tilde{G}\left(q^2,a\right)}{q^2(p-q)^2} + S_2(\mu^2) \\
 I_3(L) & = & \int\d^4q\frac{\tilde{G}\left(q^2,a\right)\tilde{G}\left((p-q)^2,a\right)}{q^2(p-q)^2}+ S_3(\mu^2).
\end{eqnarray*}
The $S_i$'s are the formally infinite counter-terms of kinematical renormalization, which ensure that \(\tilde G\) is zero at the reference impulsion $\mu$. They disappear when deriving with respect to \(L\).

Now,  $I_1$ gives the term proportional to \(a\) in \(\gamma\), and \(a\) was normalized so that $\gamma(a) = a\left(1+\mathcal{O}(a)\right)$. Otherwise $\tilde{G}$ is 0 for \(a=0\), hence $a\partial_LI_2$ (resp. $a\partial_LI_3$) starts by $a^2$ (resp. 
$a^3$), so that we have:
\begin{equation}
 \partial_LI_1(L) = -1.
\end{equation}
The Schwinger--Dyson equation is therefore written as follows:
\begin{equation}
 \gamma(a) = a\left(1-2\,\partial_L \!\int\d^4q\frac{\tilde{G}\left(q^2,a\right)}{q^2(p-q)^2} - \partial_L \!\int\d^4q\frac{\tilde{G}\left(q^2,a\right)\tilde{G}\left((p-q)^2,a\right)}{q^2(p-q)^2}\right)
\end{equation}
This equation can be mapped to the Borel plane, using the relation (\ref{primB}) to express the multiplication by \(a\). We end up with
\begin{equation} \label{SDE_B}
 \hat{\gamma}(\xi) = 1 - 2\int_0^{\xi}\d\eta\;\partial_L \!\int\d^4q\frac{\hat{G}(q^2,\eta)}{q^2(p-q)^2}
  - \int_0^{\xi}\d\eta\;\partial_L \!\int\d^4q
  \frac{\hat{G}\left(q^2,\eta\right)\star\hat{G}\left((p-q)^2,\eta\right)}{q^2(p-q)^2}.
\end{equation}
The convolution product in the last integral shall be read as
\begin{equation*}
 \hat{G}\left(q^2,\eta\right)\star\hat{G}\left((p-q)^2,\eta\right) = \int_0^{\eta}\hat{G}\left(q^2,\eta-\sigma\right)\hat{G}\left((p-q)^2,\sigma\right)\d\sigma.
\end{equation*}
Now,  using the parametrization (\ref{param_G}) of $\hat{G}$ within the Schwinger--Dyson equation (\ref{SDE_B}), we get
\begin{equation*}
 \partial_L \!\int\d^4q\frac{\hat{G}(q^2,\eta)}{q^2(p-q)^2} = \oint_{\mathcal{C}_{\eta}} \mkern -8mu
 \d\zeta\frac{f(\eta,\zeta)}{\zeta}\,\partial_L \!\int\d^4q\frac{e^{3\zeta L(q^2)}}{q^2(p-q)^2}
\end{equation*}
for the first non-trivial term in (\ref{SDE_B}). Then the derivative of the last integral evaluates to $H(3\zeta,0)=1/(1+3\zeta)$ using $L(q^2)=\ln(q^2)$. The loop integral can therefore be computed with the Mellin transform, pointing to the interesting properties of the parametrization of \(\hat G\)~(\ref{param_G}).

For the second integral the situation is essentially the same, but slightly more complicated. Using an obvious notation we have
\begin{equation*}
 \partial_L \!\int\d^4q\frac{\hat{G}\star\hat{G}}{q^2(p-q)^2} 
 = \int_0^{\eta}\d\sigma\oint_{\mathcal{C}_{\eta-\sigma}} \oint_{\mathcal{C}_{\sigma}} \mkern -8mu
 \d\zeta\d\zeta'\frac{f(\xi-\sigma,\zeta)f(\sigma,\zeta')}{\zeta\,\zeta'}\,\partial_L\!
 \int\d^4q\frac{e^{3\zeta L\left(q^2\right)}e^{3\zeta'L\left((p-q)^2\right)}}{q^2(p-q)^2}.
\end{equation*}
And, similarly to what was done in the physical plane, the derivative of the last integral can be evaluated to $H(3\zeta,3\zeta')$. Hence we end up with the Schwinger--Dyson equation in the Borel plane written in terms of $f$:
\begin{equation} \label{SDE_gamma_f}
 \hat{\gamma}(\xi) = 1 - 2\int_0^{\xi}\!\d\eta\oint_{\mathcal{C}_{\eta}}\mkern -8mu \d\zeta\frac{f(\eta,\zeta)}{\zeta(1+3\zeta)} 
 - \int_0^{\xi}\!\d\eta\int_0^{\eta}\!\d\sigma
 \oint_{\mathcal{C}_{\eta-\sigma}}\mkern - 17 mu \d\zeta\frac{f(\eta-\sigma,\zeta)}{\zeta}
 \oint_{\mathcal{C}_{\sigma}}\mkern -8mu \d\zeta'\frac{f(\sigma,\zeta')}{\zeta'}H(3\zeta,3\zeta').
\end{equation}
In these expressions, care must be taken that the Mellin transform \(H\) is not holomorphic, but meromorphic: when trying to use these formulas for the analytic continuation of \(\hat\gamma\), the different contours should not go past the poles of \(H\). In a sense, the use of the Mellin transform is more natural in this setting than in the perturbative computations.  In the perturbative computation, the Mellin transform is but a collecting device for all the integrals with different powers of the logarithms of the impulsions, while here the representation of the propagator as a combination of general powers of the squared impulsion makes its introduction unavoidable.

\subsection{Back to the perturbative computation} \label{backto}

We would like to link the Borel plane computation and the ones made in our previous work~\cite{BeCl13}.
Let us show that the perturbative study made using the formal series $A$ and $B$ is equivalent to a well-defined computation in the Borel plane. Let $f$ and $g$ be two functions of 
the structure constant $a$ involving a formal series $C$:
\begin{eqnarray*}
 f(a) & = & a^n + a^m C \\
 g(a) & = & a^p + a^q C.
\end{eqnarray*}
To simplify notations, we only take one power of \(a\) for each possible term, but the computations of~\cite{BeCl13} involve sums of such terms with varying exponents \(n\), \(m\), \(p\) 
and~\(q\), and \(C\) can represent either of the symbols \(A\) or~\(B\). \(C\) is encoding the asymptotic behavior of the functions, or equivalently a singularity of the Borel transform:
\begin{equation}
 C = \sum C_n a^n \qquad \frac{C_{n+1}}{C_n}=\alpha n -\beta
\end{equation}
with $\alpha\neq0$. This is a formal series but is Borel summable. Without loss of generality, we can assume $\alpha=1$ since we can make an expansion in $\tilde{a}=a/\alpha$. This is nothing 
but mapping the singularity of the Borel transform to $\xi=1$. When doing our perturbative analysis, we assumed that the product of the functions such as $f$ and $g$ was given by
\begin{equation} \label{prod_utile}
 f(a)g(a) = a^{n+p} + \left[a^{m+p}+a^{q+n}\right]C.
\end{equation}
We will check that this is coherent with the map into the Borel plane, that is, compute $\widehat{fg}$ and $\hat{f}\star\hat{g}$ and check that they coincide in the right limits. First, the Borel 
transform of the formal series $C$ is
\begin{equation}
 \hat{C} = \sum\frac{C_{n+1}}{n!}\xi^n = \sum\hat{C}_n\xi^n.
\end{equation}
Thus we get the recurrence relation for the $\hat{C}_n$ coefficients:
\begin{equation} \label{rec_aB}
 \frac{\hat{C}_n}{\hat{C}_{n-1}} = 1-\frac{\beta}{n}.
\end{equation}
The most natural $\hat{C}_n$ coefficients satisfying the above recurrence relations are
\begin{equation} \label{aB1}
 \hat{C}_n = c\prod_{i=1}^n\left(1-\frac{\beta}{i}\right).
\end{equation}
However, this can only be the right form for the $\hat{C}_n$'s if $\beta\notin\mathbb{N}$. Indeed, if $\beta\in\mathbb{N}$, we would have $\hat{C}_n=0$, for large enough \(n\). Since~(\ref{rec_aB}) has 
to be asymptotically true (it encodes the asymptotic behavior of $f$ and $g$), for $\beta\in\mathbb{N}$, we must take a product beginning at \(\beta + 1\) in the formula for the $\hat{C}_n$'s. For generic \(\beta\), we have an explicit formula for $\hat{C}$:
\begin{equation}\label{hatA}
 \hat{C} = (1-\xi)^{\beta-1}.
\end{equation}
Indeed, this has the right Taylor expansion around $0$. Otherwise, we could use the differential equation satisfied formally by \(C\), Eq.~(\ref{relation_symboles}), convert it to a differential equation for \(\hat C\) and see that equation~(\ref{hatA}) gives its solution up to a factor.  Then, by induction, it is easy to prove
\begin{equation}
 \widehat{a^n C} \simall{\xi}{1} \frac{(-1)^n}{(\beta)_n}(1-\xi)^{\beta+n-1}
\end{equation}
since multiplication by \(a\) corresponds to taking the primitive of the Borel transform. Here $(x)_n$  is the Pochhammer symbol defined by
\begin{equation*}
 (x)_n = \frac{\Gamma(x+n)}{\Gamma(x)} = x(x+1)\cdots (x+n-1).
\end{equation*}
In the case where \(\beta\) is not an integer, we therefore have the following equivalence relations
\begin{subequations}
 \begin{eqnarray} 
  \hat{f}(\xi) & \simall{\xi}{0} & \frac{\xi^{n-1}}{(n-1)!} \label{f0N}, \\
  \hat{f}(\xi) & \simall{\xi}{1} & \frac{(-1)^m}{(\beta)_m}(1-\xi)^{\beta+m-1}, \\
  \hat{g}(\xi) & \simall{\xi}{0} & \frac{\xi^{p-1}}{(p-1)!}, \\
  \hat{g}(\xi) & \simall{\xi}{1} & \frac{(-1)^q}{(\beta)_q}(1-\xi)^{\beta+q-1}. \label{g0N}
 \end{eqnarray}
\end{subequations}
The equivalence around \(1\) are taken modulo functions holomorphic in the neighborhood of \(1\), since any such term would either be subdominant in the asymptotic behavior of the coefficients \(f_n\) or captured by a different symbol.  Even if it coincides with it in certain cases, this notion of equivalence is therefore different from the most usual one, where one neglects what is smaller in some neighborhood of the point. One way of getting rid of these holomorphic terms is to take the difference between the analytic continuation of the Borel transform by either side of \(1\). Any holomorphic function is killed, while the non-integer powers are multiplied by \(\sin(\pi\beta)/\pi\) (for convenience, the difference is divided by \(2\pi i\)). However, such an operation annihilates functions with poles, which are however important singularities: this means that there is not a unique way to look at these singular parts, but a few options which all have their own qualities.

Let us go back to the convolution product of \(\hat{f}\) and \(\hat{g}\).  First, it is trivial to check
\begin{equation}
 \hat{f}\star\hat{g}(\xi) \simall{\xi}{0} \frac{\xi^{n+p-1}}{(n+p-1)!} = \mathcal{B}(a^{n+p}).
\end{equation}
Hence the $a^{n+p}$ term of (\ref{prod_utile}) is justified: it is just the correspondence between ordinary product and the convolution product of the Borel transform.

Now, using that $\hat{f}$ and $\hat{g}$ have only one singularity in $\xi=1$, we have
\begin{equation*}
 \hat{f}\star\hat{g}(\xi) \simall{\xi}{1} \underbrace{\int_0^\frac12 \hat{f}(t)\hat{g}(\xi-t)\d t}_{I_1(\xi)} + \underbrace{\int_\frac12^\xi\hat{f}(t)\hat{g}(\xi-t)\d t}_{I_2(\xi)}.
\end{equation*}
Let us start by $I_1$.
\begin{equation*}
 I_1(\xi) \simall{\xi}{1} \frac{(-1)^{q}}{(n-1)!(\beta)_q}\int_0^\frac12 t^{n-1}(1-\xi+t)^{\beta+q-1}\d t
\end{equation*}
Performing $n-1$ integrations by parts to get rid of the $t^{n-1}$ in the integrand and taking care of the combinatorial factors we end up with
\begin{equation} \label{I2betaNotInN}
 I_1(\xi) \simall{\xi}{1}\frac{(-1)^{q+n}}{(\beta)_{q+n}}(1-\xi)^{\beta+q+n-1} = \mathcal{B}(a^{q+n}C)
\end{equation}
The contributions from the other end point are holomorphic for \(\xi\) in the neighborhood of \(1\) and are therefore negligible.
For $I_2$ we have
\begin{equation*}
 I_2(\xi) \simall{\xi}{1} \frac{(-1)^m}{(p-1)!(\beta)_m}\int_{\frac12}^\xi (1-t)^{\beta+m-1}(\xi-t)^{p-1}\d t.
\end{equation*}
Using the transformation $x=\xi-t$ this integral becomes an integral similar to $I_1$, and similar integrations by parts give us
\begin{equation} \label{I1betaNotInN}
 I_2(\xi) \simall{\xi}{1}\frac{(-1)^{p+m}}{(\beta)_{p+m}}(1-\xi)^{\beta+p+m-1} = \mathcal{B}(a^{p+m}C).
\end{equation}
Hence, (\ref{I2betaNotInN}) and (\ref{I1betaNotInN}) justify the $\left[a^{m+p}+a^{q+n}\right]C$ term in (\ref{prod_utile}) for $\beta\notin\mathbb{N}$ through the correspondence between the asymptotic behavior of the perturbative series and the singularities of the Borel transform.

For $\beta\in\mathbb{N}^*$ we have to take another form for the $\hat{C}_n$'s. We will take
\begin{equation}
 \hat{C}_n = \frac{1}{n(n-1)\cdots(n-\beta+1)}
\end{equation}
and start the sum within $\hat{C}$ at $n=\beta$. Then:
\begin{eqnarray}
 \hat{C}(\xi) & = & \sum_{n\geq\beta}\underbrace{\int^{\xi}\ldots \int_0}_{\beta\text{ times}}t^{n-\beta}\d t \nonumber \\
              & = & \int^{\xi}\ldots \int_0\frac{\d t}{1-t} \nonumber \\
              & \simall{\xi}{1} & \frac{(-1)^\beta}{(\beta-1)!} (1-\xi)^{\beta-1}\ln(1-\xi).
\end{eqnarray}
Then, by induction, it is easy to prove
\begin{equation}
 \widehat{a^nA} \simall{\xi}{1}\frac{(-1)^{n+\beta}}{(\beta+n-1)!}(1-\xi)^{\beta+n-1}\ln(1-\xi).
\end{equation}
For $\beta=0$ no integration has to be performed when computing $\hat{C}$ and hence $\hat{C}(\xi) \simall{\xi}{1}(1-\xi)^{-1}$. Nevertheless, the above formula includes the case $\beta=0$. The 
equivalence relations (\ref{f0N})--(\ref{g0N}) become now
\begin{subequations}
 \begin{eqnarray}
  \hat{f}(\xi) & \simall{\xi}{0} & \frac{\xi^{n-1}}{(n-1)!}, \\
  \hat{f}(\xi) & \simall{\xi}{1} & \frac{(-1)^{m+\beta}}{(\beta+m-1)!}(1-\xi)^{\beta+m-1}\ln(1-\xi), \\
  \hat{g}(\xi) & \simall{\xi}{0} & \frac{\xi^{p-1}}{(p-1)!}, \\
  \hat{g}(\xi) & \simall{\xi}{1} & \frac{(-1)^{q+\beta}}{(\beta+q-1)!}(1-\xi)^{\beta+q-1}\ln(1-\xi). \\
 \end{eqnarray}
\end{subequations}
Following the same strategy than for the case $\beta\notin\mathbb{N}$ we find that the combinatorial factors nicely combine such that
\begin{eqnarray}
 \hat{f}\star\hat{g}(\xi) & \simall{\xi}{1} & - \frac{(\xi -1)^{\beta+q+n-1}}{(\beta+q+n-1)!}\ln(1-\xi) - \frac{(\xi -1)^{\beta+m+p-1}}{(\beta+m+p-1)!}\ln(1-\xi) \nonumber \\
                          & = & \mathcal{B}(a^{q+n}C) + \mathcal{B}(a^{m+p}C).
\end{eqnarray}
Thus our perturbative computations are strictly equivalent to computations around the singularities of the Borel transform. Here we see that the Borel transform approach to the Schwinger--Dyson equation 
allows for a more natural interpretation of our results. 

Moreover, let us notice that neither $A$ nor $B$ can appear alone in
$\gamma$. The lowest order terms are $aA$ and $aB$. Hence they correspond
in the Borel plane to singularities at $\xi=\pm1/3$ and 
\begin{eqnarray*}
 \widehat{aA} & \simall{\xi}{-1/3} & \left(\xi+\frac{1}{3}\right)^{-5/3} \\
 \widehat{aB} & \simall{\xi}{1/3} & \ln\left(\xi-\frac{1}{3}\right)
\end{eqnarray*}
as stated in \cite{BeCl13}.

In fact, these computations are but the first steps in a general approach to the singularities of the Borel transform initiated some time ago by Jean \'Ecalle, the Alien calculus~\cite{Ecalle81}, an introduction of which can be found in~\cite{Sa14}. In our case, it just means that we extract the singular part of a function around \(\xi\) by taking the difference of the two analytic continuation around \(1\) and shifting to have an expansion around 0.  The coefficients of \(C\) which describe the asymptotic properties of the formal power series \(f\) and~\(g\) are therefore a description of the singularity of the Borel transforms, which can be extracted by an operator \(\Delta_1\):
\begin{eqnarray}
  \hat{f}(\xi)  \simall{\xi}{1} 
  \frac{(-1)^{m+\beta}}{(\beta+m-1)!}(1-\xi)^{\beta+m-1}\ln(1-\xi)
  &\Longrightarrow & \Delta_1 \hat{f} = - \frac{ \xi^{\beta + m -1}}{(\beta + m - 1)!},\\
  \hat{f}(\xi)  \simall{\xi}{1} \frac{(-1)^m}{(\beta)_m}(1-\xi)^{\beta+m-1}
   &\Longrightarrow &       \Delta_1 \hat{f} = \frac{-\sin(\pi\beta)}{\pi} \frac{\xi^{\beta + m -1} }{ (\beta)_m}.
\end{eqnarray}
The first line corresponds to the case where \(\beta\) is a positive integer, the second one to non-integer \(\beta\).  The computations we just
made tell us that \(\Delta_1\) is a derivation with respect to the convolution product of the functions in the Borel plane.  

The whole story is subtler, because our computation was limited to singularities of the Borel transform on the limit of the disk of convergence. In many cases, one expects that there will be singularities for any integer multiple of a given singularity.  Then the singularity of the convolution product receives contributions from the pinching of the integration contour between singularities of \(\tilde f(\eta)\) and \(\tilde g(\xi - \eta)\). However \'Ecalle has shown that, by summing the singularities of the \(2^k\) differing analytic continuations of a function along paths going above or under the \(k\) singularities between the origin and a potential singularity with suitable weights, one obtains a derivation with respect to the convolution product, that he named an {\em alien}  derivation. Such derivations can then be used to compute the relation between the sums defined by integrating the Borel transform in different sectors.

Applying an alien derivation \(\Delta_\xi\) to a system of equations for the Borel transforms, one obtains a system of equations which is linear in the alien derivatives of the indeterminate functions: for generic values of the parameter \(\xi\), the only solution of this system will be zero, and we can conclude that the solutions in the Borel plane have no new singularity at this point (it is still possible to have a singularity if \(\xi\) is the sum of the positions of other singularities). At other points, there will be a one-dimensional space of solutions, which will determine the singularities at this point up to a single scale.

For finite-order computations, it is much easier to use formal series in the physical plane, which are easily multiplied by computer algebra
systems, exactly how we have done in~\cite{BeCl13}. At this stage, alien calculus is just giving us a nice interpretation beyond formal series.

\section{Singularities of the Borel transform}

\subsection{Localization of the singularities} \label{localization}

Here we will prove that any singularity of $\hat{\gamma}$ is linked to a singularity of $H$, the Mellln transform defined in Eq.~(\ref{def_H}). 

First, we need to make an assumption on the singularities of $\hat{\gamma}$. We will assume that they are of the type studied in section~\ref{backto}.
We call such singularities algebraic and they are characterized by the exponent \(\beta\) which we call its order.  
This assumption is quite natural since the singularities studied in 
\cite{BeCl13} are indeed algebraic in this sense. For now, we will prove that any algebraic singularity of $\hat{\gamma}$ has to correspond to a singularity of $H$. Hence, if $\xi_0$ is a singularity of $\hat{\gamma}$ we will write:
\begin{equation}
 \hat{\gamma}(\xi) \underset{\xi\rightarrow\xi_0}{\sim} c(\xi-\xi_0)^{\beta}
\end{equation}
with $c$ a constant. Strictly speaking, this is not an equivalence in the usual meaning of the symbol, if $\beta$ is a non-negative integer there is a
logarithmic factor and for positive real part of \(\beta\), the difference between the two terms can be any function holomorphic in the neighborhood
of \(\xi_0\). Moreover, the derivative of \(\hat\gamma\) will be equivalent in the same sense to \(c\beta(\xi-\xi_0)^{\beta-1}\), except in the case
\(\beta=0\) where we forget the factor \(\beta\). The virtue of our definition of an 
algebraic singularity is that one has not to take care if there are logarithms or not at the singularity.

We can deduce many things from the equation (\ref{renorm_f}). First, $\forall\zeta\neq\xi_0$, the function $\xi\longrightarrow f(\xi,\zeta)$ has a
singularity at $\xi=\zeta$ of order $-4/3+2\zeta$. 
Secondly, $\forall\zeta\neq\xi_0$ $\xi\longrightarrow f(\xi,\zeta)$ has a singularity at $\xi=\xi_0$ of order $\beta$ if $\hat{\gamma}$ has a
singularity of order $\beta$ in $\xi_0$. The third 
possibility is a combination of the two other ones, with $\zeta=\xi_0$. The two exponents  \(\beta-1\) and \(-4/3+2\xi_0\) appear possible, but such a
situation requires a case-by-case study.

As a function of its second argument and for any value of \(\xi\) which is not singular for \(\hat\gamma\), the function $f(\xi,\zeta)$ has a
singularity of order $-4/3+2\xi$ at $\zeta=\xi$. Let us emphasize that the function $\zeta\longrightarrow f(\xi,\zeta)$ has for only singularities
\(0\) and \(\xi\) and is in particular regular at $\zeta=\xi_0$.

Now, let us assume that $\xi_0$ is an algebraic singularity of $\hat{\gamma}$ and that $H(3\xi_0,0)$ is not singular. Then the integral over $\eta$ of the first integral of 
(\ref{SDE_gamma_f}) does not have to cross any singularity of $H$. We can deform its integration contour, then the Jordan's lemma gives
\begin{equation*}
 \oint_{\mathcal{C}_{\eta}}\d\zeta\frac{f(\eta,\zeta)}{\zeta(1+3\zeta)} = -\text{Res}\left(\frac{f(\eta,\zeta)}{\zeta(1+3\zeta)},\zeta=-1/3\right).
\end{equation*}
$\xi_0\neq-1/3$ (since $(-1,0)$ is a singularity of $H$). Furthermore, $\eta$ is running from $0$ to $\xi$, and $\xi\rightarrow\xi_0$. Then $\zeta\longrightarrow f(\eta,\zeta)$ is regular at 
$\zeta=-1/3$. Hence,
\begin{equation} \label{contm1}
 \oint_{\mathcal{C}_{\eta}}\d\zeta\frac{f(\eta,\zeta)}{\zeta(1+3\zeta)} = f(\eta,-1/3).
\end{equation}
According to (\ref{SDE_gamma_f}), $\eta\longrightarrow f(\eta,-1/3)$ has a singularity in $\xi_0$, but that singularity is of the same order than the singularity of $\hat{\gamma}(\xi)$. Since we have assumed this 
singularity to be algebraic, $\int\d\eta f(\eta,-1/3)$ is less singular than $\hat{\gamma}(\xi)$. Hence the first integral (\ref{SDE_gamma_f}) is not
sufficient to allow a singularity of $\hat{\gamma}(\xi)$ at \(\xi_0\). However, let us notice than this construction tells us that this integral will
give a dominant contribution to the singularity at $\xi=-1/3$ of $\hat{\gamma}(\xi)$.

For the second integral, using the fact that the alien derivative is a derivative with respect to the convolution product we get a relation between the singular part of $\hat{\gamma}$ and of $f$:
\begin{equation}
 \Delta_{\xi_0}\hat{\gamma}(\xi) \sim \int_0^{\xi}\d\eta\oint_{\mathcal{C}_0}\frac{\d\zeta}{\zeta}\oint_{\mathcal{C}_{\xi_0}}\frac{\d\zeta'}{\zeta'}H(3\zeta,3\zeta')f(-,\zeta)\star\Delta_{\xi_0}f(-,\zeta').
\end{equation}
Since in this equation, we are only interested in the behavior of \(\Delta_{\xi_0} \hat\gamma(\xi)\) in the vicinity of the origin, the integration
contour for \(\zeta\) can be a fixed one around 0 and the one for \(\zeta'\) a fixed contour enlacing 0 and \(\xi_0\).  
In the last loop integral, if \(H(0,3\xi)\) is not singular for any value of \(\xi\) on the straight line from \(0\) to \(\xi_0\), the
contour can be freely deformed to one contour \(\mathcal{C}_{\xi_0}\) which does not touch $\xi_0$.  Therefore, in the convolution integral,
\(\Delta_{\xi_0}f(\xi,\zeta')\) is of order \(\beta\) for all \(\zeta'\) on the contour. Then at least two 
integrals are taken from the convolution product and the explicit integration and since the loop integrals do not modify the singularity we end up
with a singularity of order $\beta-2$. The hypothesis that \(\hat\gamma\) has a singularity of order \(\beta\) is therefore incoherent, since we have
shown that it is equal to the sum of two terms which are less singular.

In the case where \(H(0,3\xi_0)\) is singular,  this argument does not hold:  we cannot deform the contour to include $\xi_0$ without modifying the value of 
the integral. Hence, when $\xi\rightarrow\xi_0$, the contour is pinched between $\xi$ and $\xi_0$ and there is a contribution from
$f(\xi,\zeta=\xi_0)$. 

\subsection{Study of the negative singularities} \label{negative}

We will now study the behavior of $\hat{\gamma}$ near the singularities on the negative real axis. We will use the equations (\ref{renorm_f}) and (\ref{SDE_gamma_f}). First, let us notice that the 
function $\xi\longrightarrow f(\xi,\zeta)$ can be expressed near $0$ as
\begin{equation} \label{f_zero}
 f(\xi,\zeta)\simall{\xi}{0}\sum_{p=1}^{+\infty}\frac{\hat{\gamma}_p(\xi)}{(3\zeta)^p}.
\end{equation}
Indeed, using
\begin{equation*}
 \oint_{\mathcal{C}_{\xi}}\frac{e^{3\zeta L}}{(3\zeta)^p}\frac{\d\zeta}{\zeta} = \frac{L^p}{p!}
\end{equation*}
we get, when using (\ref{f_zero}) in (\ref{param_G})
\begin{equation}
 \hat{G}(\xi,L) = \sum_{p=1}^{+\infty}\hat{\gamma}_p(\xi)\frac{L^p}{p!}.
\end{equation}
And this is exactly the Borel transform of (\ref{2ptB}). Let us remark that the above expression is well defined as a formal series in $\xi$ since each \(\hat\gamma_p\) is of order \(p\) in \(\xi\). 
On the other side, in the vicinity of a singularity of $\hat\gamma$, all the $\hat{\gamma}_p$ have the same kind of singularity and the above
expression is no longer clearly convergent:  it is better to use the parameterization~(\ref{param_G}).

From equation~(\ref{contm1}), the term linear in $\hat{G}$ in (\ref{SDE_gamma_f}) will give a contribution proportional to \(f(\xi,-1/3)\) and will
make the case \(\xi_0=-1/3\) special.  
From now on, we will focus on the cases $\xi_0\neq-1/3$. To study the contribution of the term $\hat{G}\star\hat{G}$ of equation 
(\ref{SDE_gamma_f}) to a negative singularity of $\hat{\gamma}$, let us split $H(3\zeta,3\zeta')$ between a regular and a singular part:
\begin{equation} \label{H_split}
  H(3\zeta,3\zeta') = \tilde{H}_k(3\zeta,3\zeta') + \sum_{l=1}^k\biggl(\frac{P_l(3\zeta)}{3\zeta'+l} + \frac{P_l(3\zeta')}{3\zeta+l} \biggr)                                                                                                 
\end{equation}
Since the term $\tilde{H}_k$ is regular up to \(3\zeta = -k-1\), the integration contours can be deformed and it will not give dominant contributions to the singularity of $\hat{\gamma}$: it is the same analysis than the one done to localize the singularities of $\hat{\gamma}$ in the previous 
subsection. The singular term being simple rational functions, we can once again compute the integral on the pole part using the Jordan's lemma.
\begin{equation} \label{pole}
 \oint_{\mathcal{C}_{\sigma}}\d\zeta'\frac{f(\sigma,\zeta')}{\zeta'}\frac1{3\zeta'+k} = \frac{f(\sigma,-k/3)}{k}
\end{equation}
This equality is established for \(\sigma\) in the vicinity of 0, but can be extended by analytic continuation.  Similarly, the integration for a
monomial \( (3\zeta)^m\) can be easily established for \(\sigma\) in the vicinity of 0 using~(\ref{f_zero}) and extended to the whole Borel plane:
\begin{equation} \label{monom}
 \oint_{\mathcal{C}_{\sigma}}\d\zeta\frac{f(\sigma,\zeta)}{\zeta} (3\zeta)^m = \hat\gamma_m(\sigma).
\end{equation}
Now, we only want the most singular part of the quadratic in \(f\) term. This cannot come from the regular part of \(H\) and the contributions of the
poles can be written using (\ref{pole}) and~(\ref{monom}) as a sum of terms \(\hat\gamma_m\star f(-,-l/3)\).  For the singularity in \(-k/3\), the
most singular of these terms is \(\hat\gamma\star f(-,-k/3)\) and we obtain:
\begin{equation}
 \hat{\gamma}(\xi)\simall{\xi}{\xi_0}-2\int\d\eta\int\d\sigma\; \hat{\gamma}(\eta-\sigma)f(\sigma,-k/3)\frac1{k(k-1)}
\end{equation}
since the linear part of \(P_k\) is \(-x/(k-1)\).
Hence, if $\hat{\gamma}$ has a singularity of order $\beta_k$ at $\xi=-k/3$, 
then $f(\xi,-k/3)$ has to have a singularity of order $\beta_k-2$. 
We then get a relation between $c_k$, the leading coefficient of $\hat{\gamma}$ and $f_k$, the leading coefficient of $f(\xi,-k/3)$:
\begin{equation} \label{rel_c_f2_int}
 c_k = \frac{-2}{k(k-1)}\frac{f_k}{\beta_k(\beta_k-1)}.
\end{equation}
To find $\beta_k$, we are a priori in the complicated case where \(\hat\gamma\) has a singularity for the value of \(\zeta\).  However, since
the order of \(\hat\gamma\) is small enough,  the renormalization group equation (\ref{renorm_f}) at its most singular order \(\beta_k-1\) takes the simple form:
\begin{equation*}
 3 f_k = \frac{- f_k}{\beta_k-1} + \frac{6\xi_0 f_k}{\beta_k-1}
\end{equation*}
 using $\hat{\gamma}(0)=1$ and $\hat{\gamma}'(0)=-2$. Using $\xi_0=-k/3$ we get
\begin{equation}
 \beta_k = -\frac{2}{3}(k-1).
\end{equation}
Hence the relation (\ref{rel_c_f2_int}) becomes
\begin{equation}
 c_k = -\frac{9}{k(k-1)^2(2k+1)}f_k.
\end{equation}
Now, let us go back to the case $\xi_0=-1/3$. From the previous analysis, the most singular term in the Schwinger--Dyson equation~(\ref{SDE_gamma_f})
is the one linear in \(f\), so that \(f(\xi,-1/3)\) must be of order \(\beta_1-1\) and we have the following relation between the leading coefficients
around \(\xi_0\) of \(\hat\gamma\) and \(f(\xi,-1/3)\):
\begin{equation} \label{rel_coef_premier}
 \beta_1 c_1 = 2 f_1.
\end{equation}
In the renormalization group equation~(\ref{renorm_f}), the leading singularity is now of order \(\beta_1\) and its coefficient includes a contribution from $\hat{\gamma}$:
\begin{equation*}
 -3f_1 = c_1 + \frac{f_1}{\beta_1}-6\left(-1/3\right)\frac{f_1}{\beta_1}.
\end{equation*}
Then using (\ref{rel_coef_premier}) we get
\begin{equation}
 \beta_1 = -5/3, \quad \quad c_1 = -\frac{6}{5}f_1,
\end{equation}
in conformity with the result found in~\cite{BeCl13}.

\subsection{Study of the positive singularities} \label{PosSing}

For the positive singularities, the previous analysis has to be modified. Indeed, the denominators in the poles are of the form \(k - 3\zeta
-3\zeta'\) and the residues as a function of \(\zeta'\) would involve $f(\xi,k/3-\zeta)$. When performing the 
second contour integral, the relation (\ref{f_zero}) then tells us that we would have to take derivatives of $f$ with respect to its second argument, and the renormalization group equation 
(\ref{renorm_f}) implies that those derivatives are as singular as the first term, so all of them would need to be taken into account. This would make the analysis intractable in practice.

In our previous work~\cite{BeCl13}, we determined a renormalization group like equation satisfied by the contribution \(L_k\) stemming from a pole term
in the Mellin transform \(H\). We will simply translate this equation in the Borel plane. The positive pole of order \(k\) was written 
\begin{equation*}
 \frac{Q_k(xy)}{k-x-y},
\end{equation*}
with the residue written in terms of a polynomial \(Q_k\) of degree \(k\):
\begin{equation*}
 Q_k(X) = \sum_{i=1}^{k}q_{k,i}X^i.
\end{equation*}
Then the equations for the $L_k$ functions are:
\begin{equation}
 (k-2\gamma-3\gamma a\partial_a)L_k = \sum_{i=1}^kq_{k,i}\gamma_i^2.
\end{equation}
Using the rules of the Borel transform, we map this equation into the Borel plane.
\begin{equation*}
 k\hat{L}_k-2\hat{\gamma}\star\hat{L}_k-3\hat{\gamma}\star\partial_{\xi}\left(\xi\hat{L}_k\right) = \sum_{i=1}^kq_{k,i}\hat{\gamma}_i\star\hat{\gamma}_i
\end{equation*}
As in the renormalization group equation, we integrate by parts the second convolution integral, using once again $\hat{\gamma}(0)=1$ to get 
\begin{equation} \label{eqLk}
 (k-3\xi)\hat{L}_k(\xi) = 2\hat{\gamma}\star\hat{L}_k(\xi) + 3\hat{\gamma}'(\xi)\star\left(\text{Id}.\hat{L}_k\right)(\xi) 
 	+ \sum_{i=1}^k q_{k,i}\hat{\gamma}_i\star\hat{\gamma}_i
\end{equation}
where the `$.$' in the second convolution integral has to be read as the pointwise product over functions.

If \(\hat L_k\) has a singularity in a point \(\xi_0\), the Schwinger--Dyson equation implies that \(\hat\gamma\) has also a singularity, but with the
order of a primitive of \(\hat L_k\).  Now, near a singularity at $\xi=\xi_0$, let us parametrize the singularity of $\hat{\gamma}$ and 
$\hat{L}_k$ by:
\begin{eqnarray*}
 \hat{\gamma}(\xi) & \simall{\xi}{\xi_0} & \frac{c_k}{\alpha_k}\left(\xi-\xi_0\right)^{\alpha_k} \\
    \hat{L}_k(\xi) & \simall{\xi}{\xi_0} & c_k\left(\xi-\xi_0\right)^{\alpha_k-1}.
\end{eqnarray*}
Now, the question is whether equation (\ref{eqLk}) allows a singular \(\hat L_k\).  The right-hand side terms are less singular than \(\hat L_k\), so
that the only possibility is when the factor \(k-3\xi\) vanishes: we then have that \(\xi_0 = k/3\). From the recursion relation for the 
$\hat{\gamma}_i$s, it is easy to see that no $\hat{\gamma}_i\star\hat{\gamma}_i$ will contribute. Indeed, the most singular term is for $i=1$ and $\hat{\gamma}\star\hat{\gamma}$ is singular as the second primitive of $\hat{L}_k$.
The most singular terms in~(\ref{eqLk}) can therefore be written and give:
\begin{equation*}
 c_k(k-3\xi)\left(\xi-\frac{k}{3}\right)^{\alpha_k-1} = 2\frac{\hat{\gamma}(0)}{\alpha_k}c_k\left(\xi-\frac{k}{3}\right)^{\alpha_k} +
 3c_k\frac{\hat{\gamma}'(0)\xi}{\alpha_k}\left(\xi-\frac{k}{3}\right)^{\alpha_k}
\end{equation*}
Now, simplifying this relation, using $\hat{\gamma}(0)=1$ and $\hat{\gamma}'(0)=-2$ and evaluating the remaining \(\xi\) as $k/3$, we end up with the simple 
formula for $\alpha_k$. 
\begin{equation}
 \alpha_k = \frac{2}{3}(k-1).
\end{equation}
Notice that for the positive singularities, no singularity has to be treated separately. Moreover, for $k=1$, we find $\alpha_k=0$, that is, a logarithmic singularity, as we found in our 
previous work and in section \ref{backto}.

\subsection{Transcendental Content of the Borel transform}

Now, a very natural question to ask is what the number-theoretical content of $\hat{\gamma}$ near its singularities is. However, the equations for the singular parts are linear, so that these singular parts are only determined up to a global constant which will be determined by matching with numerical determination of the singularity. Therefore, whenever we speak of the number-theoretical content or the weight of a coefficient in the expansion of a singularity, we really speak of the ratio of this coefficient with respect to this global constant. 
In the study of \cite{BeCl13} the first orders were computed in the physical 
plane around the two first singularities of $\hat{\gamma}$ (i.e., around $\xi_0=\pm1/3$). It was found that the expansion of $\hat{\gamma}$ around
those poles were rational products 
of odd zeta values.

Even this simple fact was very technical to prove in the physical plane because it involved the computation of complicated series and identities among multizeta values to show the annulation of the terms of highest weight. We will see that it
is much simpler to show this result in the 
Borel plane. However, a quite striking remark made in the physical plane is that the weights of those odd zetas were lower than expected at a given order. Here, our study in the Borel plane 
allows us to put a bound on those weights that is saturated by the weights found in \cite{BeCl13}.

Throughout this subsection we will use the splitting (\ref{H_split}) and replace $H$ by the relevant $\tilde{H}_k$ or its equivalent for the positive
singularities, since the polar parts do not change the transcendental content of the equation. Moreover, 
getting rid of the polar parts allows evaluating $\tilde{H}$ (which will denote the properly subtracted \(H\) in each case) at the singular point of $H$. Now, from the renormalization group equation (\ref{renorm_f}) with $\xi$ near a singularity, we see that 
one can expand $f(\xi,\zeta)$ near a singularity:
\begin{equation} \label{f_all}
 f(\xi,\zeta)\sim \sum_{\substack{r\geq0 \\ s\geq1}} \frac{\psi_{r,s}(\xi)} {\zeta^r(\zeta-\xi_0)^s}
 \sim \sum_{\substack{r\geq0 \\ s\geq1}}\sum_{n\geq0}\frac{\psi_{r,s}^{(n)}} {\zeta^r(\zeta-\xi_0)^s} (\xi-\xi_0)^{\alpha_k+r+s-1+n}
\end{equation}
with $\psi_{r,s}^{(n)}\in\mathbb{C}$. This comes from writing the L.H.S. of (\ref{renorm_f}) as $3(\xi_0-\xi+(\zeta-\xi_0))$. The $1/\zeta^r$ terms
come from the expansion~(\ref{f_zero}) of $f(\eta,\zeta)$ with $\eta$ near $0$, which get multiplied by the singular part of \(\hat\gamma\) or \(\hat\gamma'\).
Using this in the Schwinger--Dyson equation (\ref{SDE_gamma_f}) for $\xi\rightarrow\xi_0$, we see that the  loop integral in the factor where \(f\)
is singular in \(\xi_0\) will give derivatives of $\tilde{H}$, evaluated at 
$(3\zeta,0)$ and $(3\zeta,3\xi_0)$. The other \(f\) has only to be taken in the vicinity of 0 so that the expansion~(\ref{f_zero}) can be used, and
the second contour integral will ensure that we only have to evaluate \(\tilde H_k\) together with its derivatives at the points \((0,0)\) and
\((0,3\xi_0)\). Using the classical relation
\begin{equation*}
 \ln\Gamma(z+1) = -\gamma z+\sum_{k=2}^{+\infty}\frac{(-1)^k}{k}\zeta(k)z^k,
\end{equation*}
one can rewrite the Mellin transform as:
\begin{equation} \label{H_zeta}
 H(x,y) = \frac{1}{1+x+y}\exp\left(2\sum_{k=1}^{+\infty}\frac{\zeta(2k+1)}{2k+1}\left((x+y)^{2k+1}-x^{2k+1}-y^{2k+1}\right)\right)
\end{equation}
and $\tilde{H}$ only differs by rational terms around \((0,0)\), so that its derivatives have the same transcendental content as the above expression.
When taking values around \((0,3\xi_0)\), we can use the functional relation on \(\Gamma\) and the fact that \(3\xi_0\) is an integer to show that \(\tilde H(x,k+y)\) is a rational multiple,
with rational coefficients, of \(H(x,y)\), again up to the addition of a rational fraction with rational coefficients. Therefore, in every case, the
only transcendental numbers which can appear are the odd zeta values, with a total weight which is bounded by the total number of derivatives.  
Using this information in a recurrent determination of the higher order correction to the singular behavior of \(\hat\gamma\) and all the coefficients
\(\psi_{r,s}^{(n)}\), we see that only these transcendental numbers can appear.  Hence we have proved that the expansions of $\hat{\gamma}$ around its singularities have no even zeta values, nor MultiZeta 
Values that cannot be expressed as $\mathbb{Q}$-linear combinations of products of odd zetas.

\subsection{Weight of the odd Zetas}

Now, let us try to be more specific and get a bound on the weights of the different coefficients.  To study the expansion of $\hat{\gamma}$, let us expand it around a singularity:
\begin{equation}
 \hat{\gamma}(\xi) \simall{\xi}{k/3}\sum_{p=0}^{+\infty}c_k^{(p)}(\xi-k/3)^{\alpha_k+p}.
\end{equation}
We will also need the expansion of $\hat{\gamma}$ around $0$
\begin{equation}
 \hat{\gamma}(\xi) \simall{\xi}{0} \sum_{p=0}^{+\infty} c_p\xi^p.
\end{equation}
In \cite{BeSc08} it was shown that $w(c_p)=p$, with the usual weight function defined by $w\left(\zeta(n)\right)=n,w(a.b)=w(a)+w(b), w(0)=-\infty$ and $w(a+b)=\max\{w(a),w(b)\}$.  For $p=1,2$,  the weight is $0$ and for $p=4$ the weight is only $3$, both due to the absence of $\zeta(2)$ in the expansion of $\hat{\gamma}$.
In the general cases, computations must be done with alien derivatives since the expansion around the other singularities of any
particular analytic continuation will involve terms stemming from iterated alien derivatives. This does not really change the computations, but the
present formulation becomes inexact. Therefore, we will work here only for the two first singularities of $\hat{\gamma}$ and work out explicitly the case $k=+1$.

Since $\hat L_1(\xi)$ carries the most singular contribution to $\hat{\gamma}(\xi)$ for $\xi\sim \xi_0=1/3$, it is natural to assume that it will also carry the zetas of highest weight. So let us 
expand it around this singularity:
\begin{equation}
 \hat L_1(\xi) \simall{\xi}{\xi_0} \sum_{n=0}^{\infty}L_1^{(n)}(\xi-\xi_0)^{\alpha_1+n-1}.
\end{equation}
The study of \ref{PosSing} implies that the singular term of order $\alpha_1+n-1$ in $\hat L_1$ contributes to the singular term of order $\alpha_1+n$ in $\hat{\gamma}$. Our aim is to show that other contributions of this order to \(\hat\gamma\) are of lower weight so that the weights in \(\hat L_1\) determine those in \(\hat\gamma\).
To study the weight of $L_1^{(n)}$ we use the renormalization 
group equation (\ref{eqLk}).  In the neighborhood of \( \xi_0 \), the point wise multiplication by \( \xi \) does not lower the order of the singularity, so that the most singular part of the RHS of eq.~(\ref{eqLk}) comes from the term convoluted with $\hat{\gamma}'$, with \(\hat L_1\) the singular factor.  Indeed, \(L_1\) is of order 2 at the origin, so that \(\hat L_1\) vanishes at \(\xi=0\). Multiplication by \(k - 3\xi\) in the LHS lowers the order by one so that we end up with
\begin{equation} \label{recurrence_poids}
 w(L_1^{(n)}) \leq \text{max}_{p\in[2,n+1]}\{w(c_p)+w(L_1^{(n-p+1)})\}
\end{equation}
from the term proportional to \( (\xi-\xi_0)^{\alpha_1+n} \).
Since the \(c_p\) appears in the relation between the coefficients of order differing by \(p-1\), the weight of \(L_1^{(n)} \) cannot be simply \(n\). However, a weight like \(3n/2\) allows terms which are not possible.  For example, at level \(2n\), \(\zeta(3)^n \) is the only term of weight \(3n\).

In fact, there is a way to describe exactly the terms which can appear in \(L_1^{(n)}\). 
We define a modified weight system \(W\) such that $W(\zeta(2n+1)) = 2n$. With this modified weight, \(c_p\) is of weight \(p-1\) and  equation (\ref{recurrence_poids}) shows that \(L_1^{(n)}\) is of maximal weight \(n\). In fact, since the weight of the odd zetas is even, all weights are even and additional terms can only appear for even orders.

Now, let us check that the contributions from all other terms have a smaller weight. In order to simplify notations and computations, we will extend the weight \(W\) to formal series by defining:
\begin{equation}\label{WeiSeries}
	W( \sum_{p=0}^\infty a_p \xi^p ) = \sup_p \bigl( W(a_p) - p \bigr).
\end{equation}
It is now easy to show that the weight of a convolution product is bounded by the weights of its factors:
\begin{equation} \label{WeiPro}
	W ( \hat f  \star \hat g ) \leq W(\hat f) + W(\hat g) -1.
\end{equation}
Let us remark that a negative weight implies that  the first terms in the series are zero.
We will also need a similar definition around a singularity \(\xi_0\), defining the weight function \(W_{\xi_0} \) from the weights of the expansion of a function around \(\xi_0\). Here the definition will depend on the reference exponents \(\alpha_k\).  For example, we will have that 
\begin{equation}
W_{k/3}(\hat \gamma) = \sup_p (c_k^{(p)} - p ).
\end{equation}
Using the properties of the singular part of a convolution product, we can generalize formula~(\ref{WeiPro}) to
\begin{equation}\label{WeiProS}
  W_{\xi_0} ( \hat f  \star \hat g ) \leq \max(W_{\xi_0}(\hat f) + W(\hat g)-1, W(\hat f) + W_{\xi_0}(\hat g) -1) 
\end{equation}
The hypothesis we want to prove take the simple form 
\begin{equation}
 	W_{1/3} ( \hat \gamma ) = 0.
\end{equation}

Let us suppose that this is the case. Using the weights of the convolution products, Eq.~(\ref{eqLk}) shows that \(W(\hat L_1) = -1\), and then that, with our hypothesis, \(W_{1/3}(\hat L_1) = +1\) since \(W(\hat\gamma') =0\) and \(W_{1/3}(\hat\gamma') = + 1\). The difficult part is to show that the additional terms in the Schwinger--Dyson equation~(\ref{SDE_gamma_f}) depending on the subtracted Mellin transform \(\tilde H_1\) are really subdominant. 

We will need the weight of $\hat{\gamma}_n$, which can be easily deduced from its recursive definition and the relation~(\ref{WeiPro})
\begin{equation}
 W(\hat{\gamma}_n) = 1-n.
\end{equation}
Now, using this expansion, the representations (\ref{f_zero}) and (\ref{f_all}) of $f(\xi,\zeta)$ and the splitting (\ref{H_split}) of $H$ in the Schwinger--Dyson equation 
(\ref{SDE_gamma_f})\footnote{more precisely, in the term of (\ref{SDE_gamma_f}) quadratic in $\hat{G}$ since the linear one will not 
bring any new zeta.}, we end up with
\begin{equation}
 \partial_{\xi}\hat{\gamma}(\xi) \simall{\xi}{\xi_0} \sum_{p\geq1}\sum_{r\geq0}\sum_{s\geq1} \bigl(h_r^p+\tilde{h}_s^p \bigr) \hat\gamma_p \star \psi_{r,s}
\end{equation}
with the equivalence sign meaning here up to rational terms. The quantities $h_r^p$ and $\tilde{h}_s^p$ are defined by:
\begin{subequations}
 \begin{eqnarray}
          h_r^p & := & \left.\frac{\d^p}{\d\zeta^p}\left(\sum_{i=0}^r q_i^r\frac{\d^i}{\d\zeta'^i}\tilde{H}(3\zeta,3\zeta')|_{\zeta'=0}\right)\right|_{\zeta=0} \\
  \tilde{h}_s^p & := & \left.\frac{\d^p}{\d\zeta^p}\left(\sum_{i=0}^{s-1} \tilde{q}_i^s\frac{\d^i}{\d\zeta'^i}\tilde{H}(3\zeta,3\zeta')|_{\zeta'=\xi_0}\right)\right|_{\zeta=0}
 \end{eqnarray}
\end{subequations}
with $q_i^r,\tilde{q}_i^s\in\mathbb{Q}$. \(h_r^p\) (resp.\ \(\tilde h_s^p\)) have therefore a weight bounded by \(p+r\) (resp.\ \(p+s-1\)). 

The only thing that is left to find is $W_{1/3}(\psi_{r,s})$ from the renormalization group equation~(\ref{renorm_f}).  One readily obtains that this weight is bounded by \(1-r-s\).  Using that \(s\) is bounded below by 1 and the law for the convolution products, we find that every term in the sum have weight less than or equal to 0. The weight in \(\xi_0\) of \(f(\xi,-1/3)\) is also bounded by 0 so we verify that all these terms have subdominant weights with respect to \(\hat L_1\).

The case with \(\xi_0 =-1/3\) is quite similar: the only real difference is that, due to the presence of \(f(\xi,-1/3)\) in the right-hand side of the Schwinger--Dyson equation, each successive coefficient in the expansion of \(\hat\gamma\) comes from a system of equations derived from this Schwinger--Dyson equation and the renormalization group equation for \(f\). 

These weight limits are exactly the ones observed in \cite{BeCl13}. Using the formalism of alien derivations, these results should generalize to the other singularities.

\section{Asymptotic analysis of the anomalous dimension}

\subsection{Preliminary analysis}

We will take in this section $\xi\notin\mathbb{R}$ since it has been shown earlier that the singularities of $\hat{\gamma}$ all lie on the real line. Now, let us justify that, to study the 
asymptotic behavior of $\hat{\gamma}$ as a solution of the Schwinger--Dyson equation (\ref{SDE_gamma_f}), one can drop the term quadratic in $f$. This is quite a trivial fact: the asymptotics of 
$\gamma$ (in the physical plane) was given in \cite{Be10a} by the first pole of the one-loop Mellin transform. Here, this corresponds to the pole
in $\zeta=-1/3$ in (\ref{SDE_gamma_f}), for which the integral linear in $f$ is the dominant contribution.

To justify more formally this truncation, let us write
\begin{equation*}
 H(\zeta,\zeta') = \frac{1}{1+\zeta+\zeta'}\frac{\Gamma(1-\zeta-\zeta')\Gamma(1+\zeta)\Gamma(1+\zeta')}{\Gamma(1+\zeta+\zeta')\Gamma(1-\zeta)\Gamma(1-\zeta')}.
\end{equation*}
Then, the Stirling approximation $\Gamma(1+x)\sim \sqrt{2\pi}x^{x+1/2}e^{-x}$, valid for any complex \(x\) except in the immediate vicinity of
the negative real axis, leads to
\begin{equation*}
 H(\zeta,\zeta') \sim \frac{i}{1+\zeta+\zeta'}\frac{\zeta^{2\zeta}\zeta'^{2\zeta'}}{(\zeta+\zeta')^{2\zeta+2\zeta'}}
\end{equation*}
if the imaginary parts of \(\zeta\) and \(\zeta'\) are both positive. Now, let us write $\zeta'=\alpha\zeta$. Since $\xi\notin\mathbb{R}$, we can assume that $\zeta$ and $\zeta'$ are in the same quadrant of the complex plane\footnote{this can 
be done by restraining the domain where we take $\xi$.}, therefore $\Re(\alpha)>0$. Hence we arrive to
\begin{equation*}
 H(\zeta,\zeta') = \frac{i}{1+\zeta(1+\alpha)}\left(\frac{\alpha^{\alpha}}{(1+\alpha)^{1+\alpha}}\right)^{2\zeta} .
\end{equation*}
Using the definition of complex power $z^{z'}=\left(|z|^2\right)^{z'/2}e^{iz'\text{arg}(z)}$ we end up with
\begin{equation*}
 \left|\frac{\alpha^{\alpha}}{(1+\alpha)^{1+\alpha}}\right| < 1 \Leftrightarrow \alpha_1\ln\left|\frac{\alpha}{1+\alpha}\right| - \ln|1+\alpha| - \frac{\alpha_2}{2}\left[\text{atan}\left(\frac{\alpha_2}{\alpha_1}\right) - \text{atan}\left(\frac{\alpha_2}{1+\alpha_1}\right)\right] < 0
\end{equation*}
with \(\alpha_1 = \Re(\alpha)\) and $\alpha_2=\Im(\alpha)$.
Since $\alpha_1>0$, $\alpha_1\ln\left|\frac{\alpha}{1+\alpha}\right|<0$ and since the function atan is monotonic, increasing over 
$\mathbb{R}$,
$ \frac{\alpha_2}{2}\left[\text{atan}\left(\frac{\alpha_2}{\alpha_1}\right) - \text{atan}\left(\frac{\alpha_2}{1+\alpha_1}\right)\right]>0$. Therefore, $H(\zeta,\zeta')$ is exponentially small at infinity for $\Re(\zeta)>0$. 

The conclusion of this subsection is that, in a sector with positive real and imaginary values of \(\xi\),  the term quadratic in $f$ in
the Schwinger--Dyson equation (\ref{SDE_gamma_f}) will involve an exponentially small \(H(\zeta,\zeta')\), except when one of the argument is in
the vicinity of 0.  It is therefore plausible that the contribution of this quadratic part remains subdominant and can be ignored without any
dramatic change of the asymptotic behavior of the solution 
for $\Re(\xi)>0$ and $\xi$ far enough of the real line.

\subsection{Truncated Schwinger--Dyson equation}

First, we can solve a specialization of the renormalization group equation (\ref{renorm_f}). Defining $g(\xi)=f(\xi,-1/3)$ and specializing (\ref{renorm_f}) to $\zeta=-1/3$ leads to
\begin{equation} \label{eq_g}
 -(1+3\xi)g(\xi)=\hat{\gamma}(\xi)+\int_0^{\xi}\hat{\gamma}(\xi-\eta)g(\eta)\d\eta +3\int_0^{\xi}\hat{\gamma}'(\xi-\eta)\eta g(\eta)\d\eta.
\end{equation}
This equation can be solved by adding a parameter $\lambda$ and writing $g$ as a series in this parameter.
\begin{equation*}
 g(\xi) = \sum_{n\geq0} \lambda^n g_n(\xi)|_{\lambda=1}
\end{equation*}
Then (\ref{eq_g}) gives the recurrence relations amongst the $g_n$'s.
\begin{eqnarray*}
               g_0(\xi) & = & -\frac{\hat{\gamma}(\xi)}{3\xi+1} \\
  -(1+3\xi)g_{n+1}(\xi) & = & \underbrace{\int_0^{\xi}\hat{\gamma}(\xi-\eta)g_n(\eta)\d\eta}_{=I_1^{n+1}(\xi)} + \underbrace{3\int_0^{\xi}\hat{\gamma}'(\xi-\eta)\eta g_n(\eta)\d\eta}_{=I_2^{n+1}(\xi)}
\end{eqnarray*}
Hence we can write the recurrence relations for the $I$'s as well:
\begin{align*}
 & I_1^{n+1}(\xi) = -\frac{1}{3}\int_0^{\xi}\frac{\hat{\gamma}(\xi-\eta)}{\eta+1/3}\left[I_1^{n}(\eta)+3I_2^{n}(\eta)\right]\d\eta \\
 & I_2^{n+1}(\xi) = -\int_0^{\xi}\frac{\hat{\gamma}'(\xi-\eta)}{\eta+1/3}\eta\left[I_1^{n}(\eta)+3I_2^{n}(\eta)\right]\d\eta.
\end{align*}
Now we can solve the induction for the $I$'s and thus solve (\ref{eq_g}). The solution will be a Chen integral. First, let us define two functions:
\begin{align*}
 \begin{cases}
  f_0(\xi,\eta) = - \frac{\eta}{\eta+1/3} \hat{\gamma}'(\xi-\eta)\\
  f_1(\xi,\eta) = - \frac{1}{3\eta+1}\hat{\gamma}(\xi-\eta).
 \end{cases}
\end{align*}
Moreover, let $I_n=(i_1,\ldots,i_n)$ be a string of integers, with $i_k\in\{0,1\}$. Now we define the iterated integrals
\begin{equation}
 F^{I_{n}}_{0,\xi} = \int_{0\leq x_n\leq\cdots \leq x_1\leq x_0 := \xi}\left[\prod_{k=1}^{n}f_{i_k}(x_{k-1},x_{k})\right]\hat{\gamma}(x_n)\d x_n\ldots \d x_1.
\end{equation}
Then the solution of (\ref{eq_g}) is:
\begin{equation} \label{sol_g}
 g(\xi) = \frac{-1}{3\xi+1}\left(\hat{\gamma}(\xi)+\sum_{n\geq1}\sum_{\{I_{n}\}}F^{I_{n}}_{0,\xi}\right).
\end{equation}

Now, according to our previous analysis, we can neglect the term $\hat{G}\star\hat{G}$ in the Schwinger--Dyson equation when looking for the asymptotic behavior of $\hat{\gamma}$. 
Hence, from (\ref{SDE_gamma_f}), we get the following equation for $\hat{\gamma}$:
\begin{equation} \label{eq_gamma}
 \partial_{\xi}\hat{\gamma}(\xi) = 2\oint_{\mathcal{C}_{\xi}}f(\xi,\zeta)\frac{1}{\zeta(1+3\zeta)}\d\zeta.
\end{equation}
Deforming the integration contour $\mathcal{C}_{\xi}$ to a circle of infinite radius, the loop integral vanishes, thanks to Jordan's lemma, and differs from the integral above only by the opposite of the residue at $\zeta=-1/3$. Hence, all in all, we get
\begin{equation}
 \partial_{\xi}\hat{\gamma}(\xi) = + 2g(\xi).
\end{equation}
And, with the solution (\ref{sol_g}), we have an equation for $\hat{\gamma}$.
\begin{equation} \label{SDE_tronquee}
 \partial_{\xi}\hat{\gamma}(\xi) = \frac{-2}{3\xi+1}\left(\hat{\gamma}(\xi)+\sum_{n\geq1}\sum_{\{I_{n}\}}F^{I_{n}}_{0,\xi}\right).
\end{equation}
This equation is coherent with $\hat{\gamma}(0)=1$ and $\hat{\gamma}'(0)=-2$.

Before we go further, let us emphasize that the relation $\hat{\gamma}'=2g$ can be used to justify our truncation scheme. Indeed, if we plug it into the renormalization group equation 
(\ref{renorm_f}) specialized to $\zeta=-1/3$, we end up with an integrodifferential equation for $\hat{\gamma}$. Taking the inverse Borel transform of this equation, we end up with a 
differential equation on $\gamma$:
\begin{equation*}
 \gamma = a - a\gamma + 2\gamma^2 - 3a\gamma\gamma'
\end{equation*}
which is exactly the equation for $\gamma$ found in \cite{Be10} (equation (17)), up to terms that do not contribute to the asymptotics of $\gamma$. Hence, this is a nice check that a solution 
$s(\xi)$ to (\ref{SDE_tronquee}) has the right asymptotic behavior.

Now, the equation (\ref{SDE_tronquee}) appears as a fixed-point equation. By defining a suitable metric on the space of functions, 
the integral operator could become contracting, proving the existence of a solution. Defining such a contracting metric is a non-trivial 
task that is left for further studies. Here, we will only numerically study the asymptotic behavior of the solution of (\ref{SDE_tronquee}).

\subsection{Numerical analysis}

Now, to study the solution $\hat{\gamma}$ numerically, we have to fix a $\xi$ and compute $\hat{\gamma}(\eta)$ for $\eta$ on the line between the origin of the complex plane and $\xi$ with 
$\hat{\gamma}(0)=1$ and $\hat{\gamma}'(0)=-2$ as initial data. We have to take $\xi$ big enough, i.e., big with respect to the periodicity of the singularities of $\hat{\gamma}$, that is $1/3$. 
$\xi$ should also not be too close to the real line for our analysis to not be spoiled by the singularities of $\hat{\gamma}$ that are known to lie on the real line. This is why we have done our 
computations with $\xi=40+35i$, which is not too big so that the algorithm runs in a reasonable time.

The difficulties of the numerical analysis come from the fact that we have to compute convolution integrals that are very sensitive to numerical instabilities. Therefore, standard tools do not 
work for them. We have used the Simpson's rule to get the following results from (\ref{SDE_gamma_f}) without the $\hat{G}\star\hat{G}$ term.
\begin{figure}
    \includegraphics{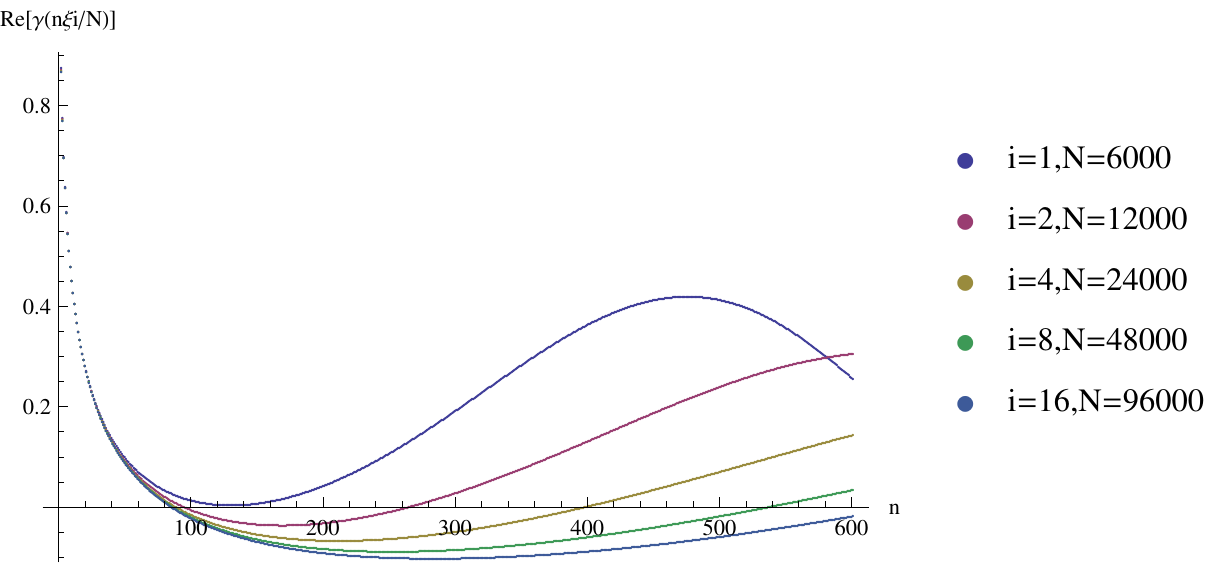}
\caption{Real part of $\hat{\gamma}$ for various precisions.} \label{ReGamma}
\end{figure}
It is clear from the above picture that a very small interval is needed in order to avoid numerical instabilities that we can see for the least precise case $i=1, N=6000$. Moreover, the minimum 
of the other curves seems to be a computational artifact since its position varies as the number of points taken increases. Although numerical methods are probably not the best way to tackle 
convolution integrals, we already see that the asymptotic behavior of the real part of $\hat{\gamma}$ seems to be a constant, eventually zero.

For the imaginary part, the same features are found, but the amplitudes are smaller (since the imaginary part of $\hat{\gamma}(0)$ is $0$), making the results harder to read.
\begin{figure}
    \includegraphics{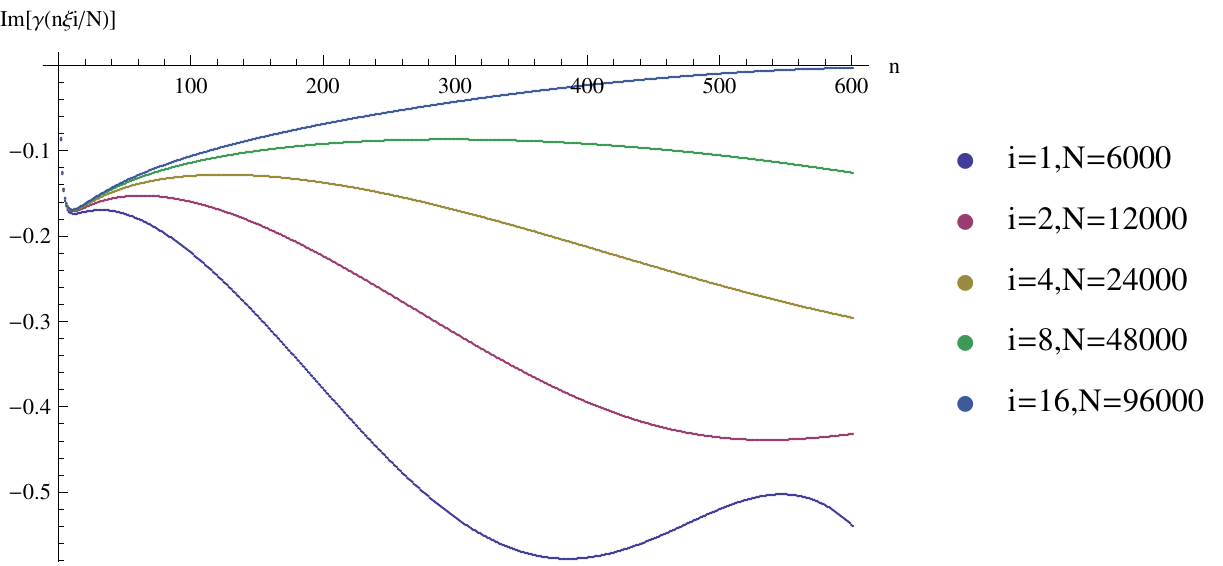}
\caption{Imaginary part of $\hat{\gamma}$ for various precisions.} \label{ImGamma}
\end{figure}
Hence, this numerical study suggests that $|\hat{\gamma}|$ is asymptotically bounded by a constant (for a non-real infinity). More precise results would require more sophisticated tools. Since we are mainly interested by analytical results, such study was not performed.

\section*{Conclusion}

We have been able to map the Schwinger--Dyson equation of the massless Wess--Zumino model into the Borel plane,  allowing an efficient study of the singularities of its 
anomalous dimension. This clarifies the role of the formal series occurring in our previous work \cite{BeCl13}.

The main results of this analysis are on the singularities of the Borel transform. We first manage to show that they all lie on the real axis, a 
result that was only conjectured so far. We also manage to find the leading order of
each singularity. Finally, we proved that only odd zetas will occur in the expansion of the anomalous 
dimension, and managed to put an upper bound on their weights. Let us notice that this bound is probably optimal since weight drops could occur only from highly non-trivial combinatorial 
mechanisms.

We have shown that the term quadratic in $\hat{G}$ in the Schwinger--Dyson equation of the massless Wess--Zumino model does not affect the asymptotic of the solution. This could be used to write 
a fixed-point equation for the asymptotic solution and to numerically study the asymptotic. That numerical studies in the Borel plane were not very conclusive, due to numerical instabilities and 
require more advanced tools.

Our main results for the number-theoretical content of $\hat{\gamma}$ have been stated for the two first singularities of the Borel transform.
Higher singularities depend on the path used to reach them. However in \'Ecalle's resurgence theory, it is shown that a suitable average of the
singularities in a point \(\xi\) reached by different paths defines a derivation, the alien derivative of index \(\xi\).  Taking the alien
derivative of the renormalization group equation and the Schwinger--Dyson equation should allow to reach complete description of the higher
singularities.
Such a study in the massless Wess--Zumino model will be the next step of our program. We aim to generalize and make more rigorous our previous results. 

The first motivation of this  work was to gain a better understanding of the results of \cite{BeCl13}. The Borel transform was particularly adapted to this task since it allows (in our case) to 
only deal with convergent series and well-defined functions. Now that our previous work is on firmer ground, we would like to extend it to more physically relevant theory, such as scalar theories, 
or even to gauge theory, maybe using tools like the corolla polynomials~\cite{KrSaSu13}. 

Finally, it would be very interesting to study the effects of higher loops corrections on the Schwinger--Dyson equation in the Borel plane. One could adapt the numerical method developed in 
\cite{Pa13}. Some issues have to be addressed before performing such a task. In particular, the numerators of the Wess--Zumino model considerably slow down the algorithm of \cite{Pa13}.

The global conclusion of this work would be that studies in the physical plane and in the Borel plane complement each others. In the physical plane, numerical computations are simpler since the product is the usual one of formal series.
On the other hand, in the Borel plane approach, one has to deal only with well-defined functions.

\bibliographystyle{unsrturl}
\bibliography{renorm}

\end{document}